\documentclass[12pt,a4paper]{article}
\usepackage{cite}

\usepackage{a4wide}
\usepackage{epsfig}
\def\be{\begin{equation}}
\def\ee{\end{equation}}

\def\be{\begin{equation}}
\def\ee{\end{equation}}
\newcommand{\ba}{\begin{array}{c}}  
\newcommand{\bad}{\begin{array}{ccc}}
\newcommand{\bea}{\begin{eqnarray}}
\newcommand{\eea}{\end{eqnarray}}
\newcommand{\ea}{\end{array}}

\usepackage{a4,graphicx}
\usepackage{amsmath}
\usepackage{psfrag}
\usepackage{rotating}
\input{standard.def}
\input{math.def}
\input{physics.def}
\input{willi.def}
\input{spezial.def}

\begin{document} 
\thispagestyle{empty}
\renewcommand{\thefootnote}{\fnsymbol{footnote}}
\setcounter{footnote}{1}
\rightline{DO-TH 03/06}
\vspace*{3mm}
\centerline{\Large \bf Possible test for $CPT$ invariance with correlated neutral}
\vspace*{3.6mm}
\centerline{\Large\bf $B$ decays}
\vspace*{7mm}
\begin{center}
{\large\bf {K.R.S. Balaji\footnote{\texttt{balaji@hep.physics.mcgill.ca}}}, 
{Wilfried Horn\footnote{\texttt{Wilfried.Horn@udo.edu}}},}
{\large\bf{ and E.A. Paschos\footnote{\texttt{paschos@physik.uni-dortmund.de}}}}\\

\medskip
\vspace*{5mm}
{\em Institut f\"ur Theoretische Physik, Universit\"at Dortmund,}\\
{\em Otto-Hahn-Stra{\ss}e 4, 44221 Dortmund, Germany}\\[3mm]
\end{center}
\noindent
\vspace*{18mm}

\begin{abstract} 
We study breakdown of $CPT$ symmetry which can occur in the decay process 
$B \bar B \to l^\pm X^\mp f$ with $f$ being a $CP$ eigenstate. 
In this process, the standard model expectations 
for time ordered semi-leptonic and hadronic events, i.e. which of the
two decays takes place first,
can be altered in the case that there
is a violation of the $CPT$ symmetry. To illustrate this possibility,
we identify and study several time integrated observables. We
find that an experiment with $10^{9}$ $B\bar B$ 
pairs, has the capability for improving the bound on $CPT$ violating 
parameter or perhaps observe $CPT$ violation.
\end{abstract} 


\renewcommand{\thefootnote}{\arabic{footnote}}
\setcounter{footnote}{0}

\newpage 
 
\section{Introduction} 

 In recent years the physics of the standard model (SM) has moved to the 
investigation of flavor physics and in particular $CP$ violation in the
$B$ meson decays. In the case of the neutral $B$ decays, the mixing induced 
$CP$ violation measurements involves a good knowledge of both the mass and 
width differences for which there is an extensive literature. 

 Most of the analyses involving physics of neutral $B$ mesons 
assume the validity of $CPT$. The
$B$ meson with their special properties provide unique opportunities of 
testing the $CPT$ symmetry. Given the eventual need for an accurate 
description of the SM $CP$ parameters, and/or tests for new physics, it is 
pertinent to also account for a possible violation of $CPT$ invariance and 
its consequences. In this regard, it is noteworthy 
that the existing experimental limits on $CPT$ violation
are not very stringent and thus it becomes necessary to allow for this
possibility in measurements of $B$ decays. An argument for this was presented 
in \cite{Kobayashi:1992uj} which questions the validity of $CPT$ theorem 
for partons which are confined states. 

It is usually assumed that $CPT$ is an exact symmetry which is hard to break.
One of the first attempts to consider the (spontaneous) breaking of  
$CPT$ came from string physics \cite{strings}. For 
instance, $CPT$ may not be a good discrete symmetry in many higher 
dimensional theories of which the SM is an effective low energy 
description \cite{Kostelecky:1991ak}. In addition, the existence of mixed 
non-commutative fields could also break $CPT$ \cite{Mocioiu:2001fx}. There 
are many tests for $CPT$ symmetry and we refer to \cite{Colladay:2003tj} for 
a review on this subject.

We present in this article a study of $CPT$ violations that originate on the
mass matrix and the analysis is model independent. We express the effects in 
terms of a complex parameter $\delta$,
whose presence modifies the expressions for width and mass differences and
affects the time development of correlated $B$ states. Precise measurements
of the time development of the states are sensitive to the presence of
$CPT$ violation. However, such studies may require large number of 
events and hence
it is preferable to have integrated rates. This motivates us 
to consider correlated decays of $B$ mesons which are time ordered. The 
procedure 
presented here requires a time ordering of leptonic/hadronic events 
(which decay happens first) without 
demanding any detailed time development of states. One reason for following 
this time ordering is because it enables for a direct extraction of the width 
differences \cite{Sinha:1998aj}. The same procedure is also 
expected to be sensitive to the presence of nonzero $\delta$ values which 
can affect the width difference. We use the time ordered rates to form 
asymmetries which 
are sensitive to $CPT$ violation. We find that our results depend on the 
width difference $y = \Delta\Gamma/(2\Gamma)$ and thus for large 
$CPT$ violating parameter $\sim O(y)$, the effects of $CPT$ can become 
significant. Our analysis is more sensitive to 
$B_s$ decays where the effects of $CP$ violation can be neglected. 
We expect that the results obtained here will complement the existing
tests for $CPT$ violation, and in particular for the $B_s$ system
where there is no information available as yet. For a recent general analysis 
of $CPT$ violation in neutral mesons, and its extraction we refer 
to \cite{kos2001}. 

Briefly, the 
possibility of $CPT$ violation has been studied in $B$ decays using 
both time dependent and integrated methods. There are several analyses 
performed which in 
most cases also involve a detailed time dependent study and/or flavor tagging
\cite{Xing:1994ep,Kostelecky:1996fk,Banuls:2000ki,Dass:2000fb,Datta:2002wa}.
In comparison with other known $CPT$ studies, we find 
certain qualitative similarities as well as differences.
Following our analysis, we find that we can restrict the strength of the
$CPT$ violating parameter to about $10\%$ which is similar to an earlier
detailed estimate \cite{Kostelecky:1996fk}. We point out that if the 
strength of
$CPT$ violating parameter is below this limit $(\sim 10\%)$, then using
our approach, it is hard to pinpoint genuine $CPT$ effects. 
In general, we find a sensitivity to $CPT$ 
effects even if the width difference is vanishingly small. This we expect
due to the particular time ordering procedure which can also extract 
corrections to the width difference due to nonzero $\delta$. In general,
in the presence of nonzero width difference, the
$CPT$ violating observables can measure
both the real and imaginary parts of the $CPT$ parameter;
while on the other hand, for zero width difference, the effects are 
sensitive only to the real part of the $CPT$ violating parameter
\cite{Banuls:2000ki}.
The latter feature is similar to what we find in our present analysis
when we assume negligible $CP$ effects. On the other hand, as
observed in \cite{Dass:2000fb},
for zero width difference, the $CPT$ violating observable is
sensitive to the imaginary part of the $CPT$ violating parameter. This
is different in our case (when we neglect $CP$ violation), as
we shall show when we discuss the $B_s$ system. 

 Our paper is organized as follows. In the next section, we present the
general formalism for the time evolution of a neutral $B$ meson in the 
presence of $CPT$ violation. In section \ref{cptvrate} we define the 
observables which can extract signals for $CPT$ violation and are 
theoretically clean.  This is followed by a brief numerical analysis for a 
specific decay channel and we compare the results with the $B_d$ system.   
We finally conclude with a brief summary in section \ref{summary}.
   
\section{The time evolution}
\subsection{Basic setup and definitions}
 In this section, we review the basic formalism for the 
decay of correlated $B \bar B$ pair in the presence of $CPT$ violation
\cite{Chou:1999wn}. 
Starting with a quantum state which is a linear combination of the $B$ and 
$\bar B$ states denoted as
\begin{align}
\ket{\psi(t)}&=\psi_1(t)\ket B+\psi_2(t)\ket {\bar B}~, 
\label{matrixformalism}
\end{align}
 the time evolution of the $B\bar B$ system can be described by 
a two dimensional Schroedinger equation 
\begin{align}
i\frac{d}{dt}\vect{\psi_1(t)\\\psi_2(t)}=
{\begin{pmatrix} 
\mathcal{M}_{11}&\mathcal{M}_{12}\\\mathcal{M}_{21}&\mathcal{M}_{22}\end
{pmatrix}}\vect{\psi_1(t)\\\psi_2(t)}.
\label{matrix1}
\end{align}
The mixing matrix is non-Hermitian and can be written as 
\be
\mathcal{M}=M-\frac{i}{2}\Gamma~,
\label{matrix}
\ee
with $M$ and $\Gamma$ being Hermitian $2\times 2$ matrices 
$\left(M=M^\dagger,\Gamma=\Gamma^\dagger\right)$. Invariance under $CPT$ 
gives
\be
\mathcal{M}_{11}=\mathcal{M}_{22} \Rightarrow M_{11}=M_{22} 
\quad\textrm{ and }\quad  \Gamma_{11}=\Gamma_{22}~,
\ee
while there is no constraint on the off-diagonal elements. As a result, to 
test $CPT$ invariance one needs to parametrize the difference of the 
diagonal elements of the mixing matrix. A sensible parameterization
has to be invariant according to a re-phasing of the meson
states \cite{Paschos:1989ur}. In particular, when a re-phasing of the type 
$\ket{{B}^\prime}\to e^{i\gamma} \ket{B}$ is done, the anti-meson state 
is altered according to 
$\ket{\overline{B}^\prime}\to e^{-i \gamma} \ket{\overline{B}}$. 
Following this, we have in the matrix elements 
\begin{align}
\mathcal{M}_{11} \to \mathcal{M}_{11}~;~  \mathcal{M}_{22} 
\to \mathcal{M}_{22}~;~
\mathcal{M}_{12} \to e^{-2i\gamma} \mathcal{M}_{12} ~;~\mathcal{M}_{21} 
\to e^{2i \gamma} \mathcal{M}_{21}~.
\end{align}
This means the diagonal elements in (\ref{matrix}) remain the same 
after re-phasing while the product of the off-diagonal elements are re-phase 
invariant. Besides, the eigenvalues of (\ref{matrix}) are re-phase
invariant. This leads to several possibilities to parameterize $CPT$ 
violation and in this analysis, we choose a parameter
\begin{equation}
\delta=\frac{\mathcal{M}_{22}-\mathcal{M}_{11}}{\sqrt{\mathcal{M}_{12}
\mathcal{M}_{21}}}~. 
\label{delta}
\end{equation}
In addition, there are other possible parameterizations which 
are rephase invariant and have been used for $CPT$ studies. Among them
we refer to parameterization in references 
\cite{Kobayashi:1992uj,cptkos,lavoura}.

In the presence of $\delta \neq 0$, the width and mass differences of the 
two $B$ states are obtained by calculating the eigenvalues of 
(\ref{matrix}). We obtain these to be

\bea
\lambda_1&=& \mathcal{M}_{11}+\sqrt{\mathcal{M}_{12}\mathcal{M}_{21}} 
\left(\sqrt{1+\frac{\delta^2}{4}}+\frac{\delta}{2}\right)~,\nonumber\\
\lambda_2&=& \mathcal{M}_{22}-\sqrt{\mathcal{M}_{12}\mathcal{M}_{21}} 
\left(\sqrt{1+\frac{\delta^2}{4}}+\frac{\delta}{2}\right )~.
\label{cptev}
\eea
We define $\mathcal{M}_{ij} = m_{ij} - i\Gamma_{ij}/2$ and 
\be
\lambda_1-\lambda_2
= -2 \sqrt{\mathcal{M}_{12}\mathcal{M}_{21}} \sqrt{\frac{\delta^2}{4}+1}=
\Delta m -\frac{i}{2}\Delta \Gamma~.
\label{egv}
\ee
Therefore upon equating the real and imaginary parts in (\ref{egv}) 
results in the width difference and mass difference
\bea
\Delta \Gamma &\equiv \Gamma_1-\Gamma_2 &= -4\abs{m_{12}} \textrm{Im}
\sqrt{1-\frac{1}{4}\frac{\abs{\Gamma_{12}}^2}{\abs{m_{12}}^2}
-i \textrm{Re} \frac{\Gamma_{12}}{m_{12}}} 
\sqrt{\frac{\delta^2}{4}+1}~,\nonumber\\
\Delta m&\equiv m_1-m_2&=2\abs{m_{12}} \textrm{Re}\sqrt{1-\frac{1}{4}
\frac{\abs{\Gamma_{12}}^2}{\abs{m_{12}}^2}
-i \textrm{Re} \frac{\Gamma_{12}}{m_{12}}} \sqrt{\frac{\delta^2}{4}+1}~.
\label{cptwd}
\eea
Furthermore, from the eigenvalue equation,
\bea
\mathcal{M}_{11}p_{1,2}+\mathcal{M}_{12}q_{1,2}&=&\lambda_{1,2}p_{1,2}
~,\nonumber\\
\mathcal{M}_{21}p_{1,2}+\mathcal{M}_{22}q_{1,2}&=&\lambda_{1,2}q_{1,2}~,
\label{egnv}
\eea
we obtain the ratios
\bea
\frac{q_1}{p_1}&=&\sqrt{\frac{\mathcal{M}_{21}}{\mathcal{M}_{12}}}\left[
\sqrt{1+\frac{\delta^2}{4}}+\frac{\delta}{2}\right]~,\nonumber\\
\frac{q_2}{p_2}&=&\sqrt{\frac{\mathcal{M}_{21}}{\mathcal{M}_{12}}}\left[
\sqrt{1+\frac{\delta^2}{4}}-\frac{\delta}{2}\right]~. 
\label{q2p2}
\eea
Introducing the mass eigenstates for the $B$ mesons as
\bea
\ket{B_{1}}&=&\frac{1}{\sqrt{p_1^2+q_1^2}}
\left[p_1\ket{B}+q_1\ket{\overline{B}}\right]\label{bl}~,\nonumber\\
\ket{B_{2}}&=&\frac{1}{\sqrt{p_2^2+q_2^2}}
\left[p_2\ket{B}+q_2\ket{\bar{B}}\right]~, \label{bh}
\eea
the time development of $B$ and $\bar B$ is evaluated to be
\bea
\ket{B(t)}&=&\frac{1}{1+\omega} \left[f_+ \ket{B}+\kappa_1 f_- 
\ket{\overline{B}}\right] \label{cpt01}~,\nonumber\\
\ket{\bar{B}(t)}&=&\frac{1}{1+\bar \omega}\left(\bar f_+ 
\ket{\overline{B}} +\bar \kappa_1 f_-\ket{B}\right)~,
\label{cpt02}
\eea
with
\bea
f_+ &=&e^{-i\lambda_1 t}+\omega e^{-i\lambda_2 t}~;~ 
\bar f_+=e^{-i\lambda_1 t}+\frac{1}{\omega} e^{-i\lambda 2 t}~,\nonumber\\
f_-&=&e^{-i\lambda_1 t}-e^{-i\lambda_2 t}~;~ 
\kappa_1=\frac{q_1}{p_1}=\frac{1}{\bar \kappa_1} ~;~
\omega=\frac{p_2q_1}{q_2p_1}~\mbox{and}~\bar\omega = \omega^*~.
\label{notations}
\eea

The information on $CPT$ violation is encoded in the complex parameter 
\be
\omega =\frac{p_2q_1}{q_2p_1}
=1+\delta+\frac{\delta^2}{2} + ~\mbox{higher order terms in}~\delta~.
\label{omega}
\ee
Given that $\delta$ is re-phase invariant, $\omega$ is also re-phase invariant 
and deviates from unity in case of $CPT$ violation.

\subsection{The decay channel $B \bar B\to l^\pm X^\mp f(\bar f)$}
In the previous section we described the formalism for $CPT$ violation that
occurs in the mass matrix. We apply it to the decays of a correlated 
$B\bar B$ pair where one of them decays semi-leptonically while the 
other decays hadronically, i.e., 
$B \bar B\to l^\pm X^\mp f(\bar f)$. The amplitude for a 
neutral $B$ meson decaying into a final state with a lepton $l^+X^-$ at time
$t_0$ and the $\bar B$ decaying in to a final state $f$ at time $t$ can be 
expressed as
\bea
A[l^+({t_0}),\bar f({t})]&=&\braket{X^- l^+}{B(t_0)}\braket{f}
{\overline{B}(t)}+ C 
\braket{X^- l^+}{\overline{B}(t_0)}\braket{f}{B(t)}~,\nonumber\\
\label{lpf}
\eea
with $C$ denoting the charge conjugation of the $B\bar B$ pair.
 The individual decay amplitudes for the hadronic channels are
\bea
\braket{f}{B}&=&A_1 e^{i\Phi_1} e^{i\eta_1}~;~
\braket{f}{\overline{B}}=A_2 e^{i\Phi_2} e^{i\eta_2}~,\nonumber\\
\braket{\bar f}{B}&=&A_2 e^{-i\Phi_2} e^{i\eta_2}~;~
\braket{\bar f}{\overline{B}}=A_1 e^{-i\Phi_1} e^{i\eta_1}~.
\label{def1}
\eea
 In (\ref{def1}), the $A_i$ denote the absolute values for the amplitudes
and $\eta_i$ and $\phi_i$ are the strong and weak phases respectively. 
Using this, along with the notations  
\be
r=\frac{A_2}{A_1}~;~\Phi=\Phi_2-\Phi_1-2\phi_m~;~
\eta =\eta_2-\eta_1 ~;~\xi=\abs{\xi}e^{2i\phi_m} =
\frac{e^{2i\phi_m}}{\sqrt{1+ {\textrm{Re}[\delta]}}}~,
\ee
with the phase due to mixing denoted by $\phi_m$, the amplitudes are now:

\bea
A[l^+f]
&=&\frac{\omega}{(1+\omega)^2}\abs{\xi}FA_1 e^{i\Phi_1}e^{i\eta_1}
e^{2i\phi_m}\nonumber\\
&&\left\{\left[f_+(t_0)f_-(t)-f_-(t_0)f_+(t)\right]
+\frac{r}{\abs{\xi}}e^{i\Phi}e^{i\eta}\left[f_+(t_0)\bar 
f_+(t)-f_-(t_0)f_-(t)\right]
\right\}~,\nonumber\\
A[l^+\bar f]
&=&\frac{\omega}{(1+\omega)^2} \bar F A_1 e^{-i\Phi_1}e^{i\eta_1}\nonumber\\
&&\left\{\left[f_+(t_0)\bar f_+(t)- f_-(t_0)f_-(t)\right]
+ r\abs{\xi}e^{-i\Phi}e^{i\eta}\left[f_+(t_0) f_-(t)-f_-(t_0)f_+(t)\right]
\right\}~,\nonumber\\
A[l^-f]
&=&\frac{\omega}{(1+\omega)^2} \bar F A_1 e^{i\Phi_1}e^{i\eta_1}\nonumber\\
&&\left\{{\left[f_-(t_0)f_-(t)-
\bar f_+(t_0)f_+(t)\right]}
+\frac{r}{\abs{\xi}}e^{i\Phi}e^{i\eta}{\left[f_-(t_0)\bar f_+(t)-
\bar f_+(t_0)f_-(t)\right]}
\right\}~,\nonumber\\
A[l^-\bar f]
&=&\frac{\omega}{(1+\omega)^2} \bar F A_1 \frac{1}{\abs{\xi}} 
e^{-i\phi_m}e^{-i\Phi_1}e^{i\eta_1}\nonumber\\
&&\left\{\left[f_-(t_0)\bar f_+(t)-\bar f_+(t_0)f_-(t)\right]
+r\abs{\xi}e^{-i\Phi}e^{i\eta}\left[f_-(t_0) f_-(t)- \bar f_+(t_0)f_+(t)
\right]\right\}~.\nonumber\\
\label{amp}
\eea
 In (\ref{amp}), $F$ denotes the amplitude for the decay $B\to l^+ X^-$ and the
corresponding amplitude for the $CP$ conjugated process is denoted by 
$\bar F$.  The decay to a lepton is measured at $t_0$ while $t$
is the time when the decay to state $f$ occurred. An explicitly
time dependence in the l.h.s. of (\ref{amp}) is assumed. 
The $B\bar B$ pair is 
taken to be the one produced at a $\Upsilon$ resonance and hence, the
$B\bar B$ charge parity, 
$C=-1$ in (\ref{lpf}). In addition, the state $f$ is chosen to be a $CP$ odd
eigenstate. This sets $\eta = 0$ and $r=-1$.

\section{The $CPT$ violating rate}
\label{cptvrate}

Following the above construction,
for the decay process $B \bar B \to l^\pm X^\mp f$ we define two time 
correlated observables: (i) $R_S$ denotes the number of events in 
which the hadronic decay 
precedes the semi-leptonic one (which in our context the time ordering is
$t < t_0$) and (ii) similarly, we define the number of events where the 
semi-leptonic decay precedes the hadronic decay denoted by $R_L$. To 
illustrate, if we choose positively charged leptons, we have the rates
\be
R_S\left[l^+f\right]=\lint_0^\infty dt_0 \lint_0^{t_0} dt \abs{A[l^+f]}^2~;
~
R_L\left[l^+f\right]=\lint_0^\infty dt_0 \lint_{t_0}^{\infty} dt 
\abs{A[l^+f]}^2~.
\label{rsrl}
\ee
 Using (\ref{rsrl}), for CP eigenstates \footnote{This choice
sets $f= \bar f$ in (\ref{lpf}).} we can define the following 
ratios
\begin{eqnarray}
R^\pm_1&=&\frac{R_S\left[l^+ f\right]\pm R_L\left[l^- f\right]}
{R\left[l^+ f\right]+R\left[l^- f\right]}~,\nonumber\\
R_2&=&\frac{R_L\left[l^+ f\right]- R_L\left[l^- f\right]}
{R\left[l^+ f\right]+R\left[l^- f\right]}
~,\nonumber\\
R_3&=&\frac{R_S\left[l^+ f\right]- R_S\left[l^- f\right]}
{R\left[l^+ f\right]+R\left[l^- f\right]}
~.
\label{cpt0}
\end{eqnarray}
where the rates without the subscripts $L$ or $S$ denote total time integrated
rates without any time ordering. The exact expressions for the above ratios
can be obtained by substituting the expressions for the rates given in the
appendix (section \ref{appendix}) of this paper. Due to the extensive nature 
of the analytic expressions involved, we avoid from presenting them in the
main text as they are also not illustrative. 
\subsection{Numerical analysis}
\label{numerical}
\begin{figure}[h]
 \begin{minipage}{0.42\textwidth}
    \begin{center}
      \psfrag{x}{$R_1^-$}
      \psfrag{y}{$\textrm{Re}\left[\delta\right]$}
\includegraphics[width=\columnwidth]{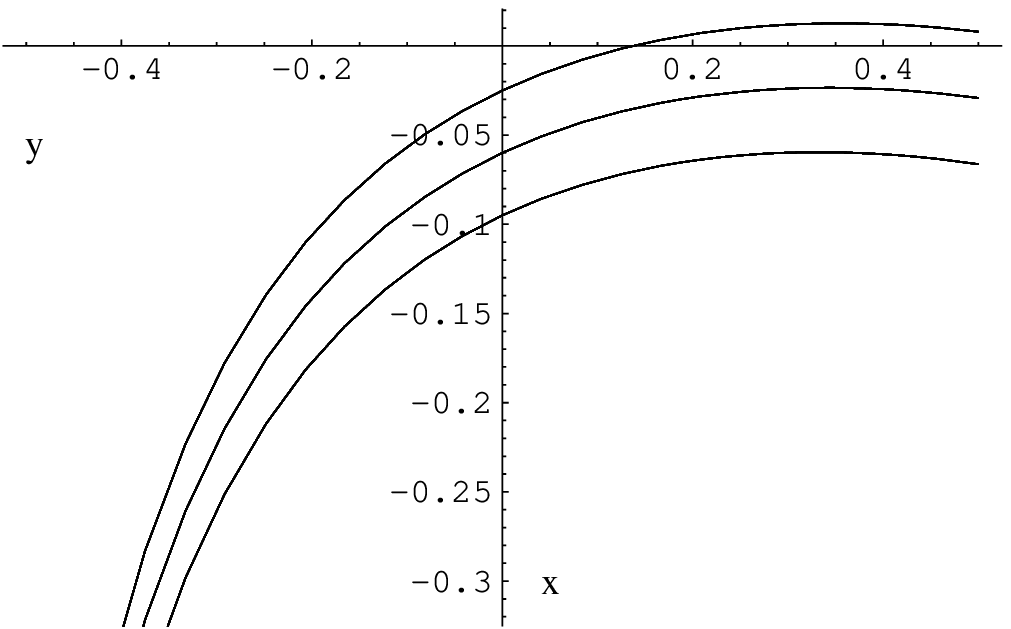}  
      \caption{Ratio $R_1^-$ versus $\textrm{Re}[\delta]$ for 
the $B_s$ system. The three curves correspond to 
$|y|= 0.05$ (upper line), $|y|= 0.12$ (middle line) and $|y|=0.19$ 
(lower line).}
      \label{bsR1MinusDelta_2DPlot_RealPart_Streifen}
    \end{center}
  \end{minipage}
  \begin{minipage}{0.68\textwidth}
    \begin{center}
      \psfrag{z}{$R_2$}
       \psfrag{x}{$\textrm{Im}\left[\delta\right]$}
        \psfrag{y}{$\textrm{Re}\left[\delta\right]$}
 \includegraphics[width=\columnwidth]{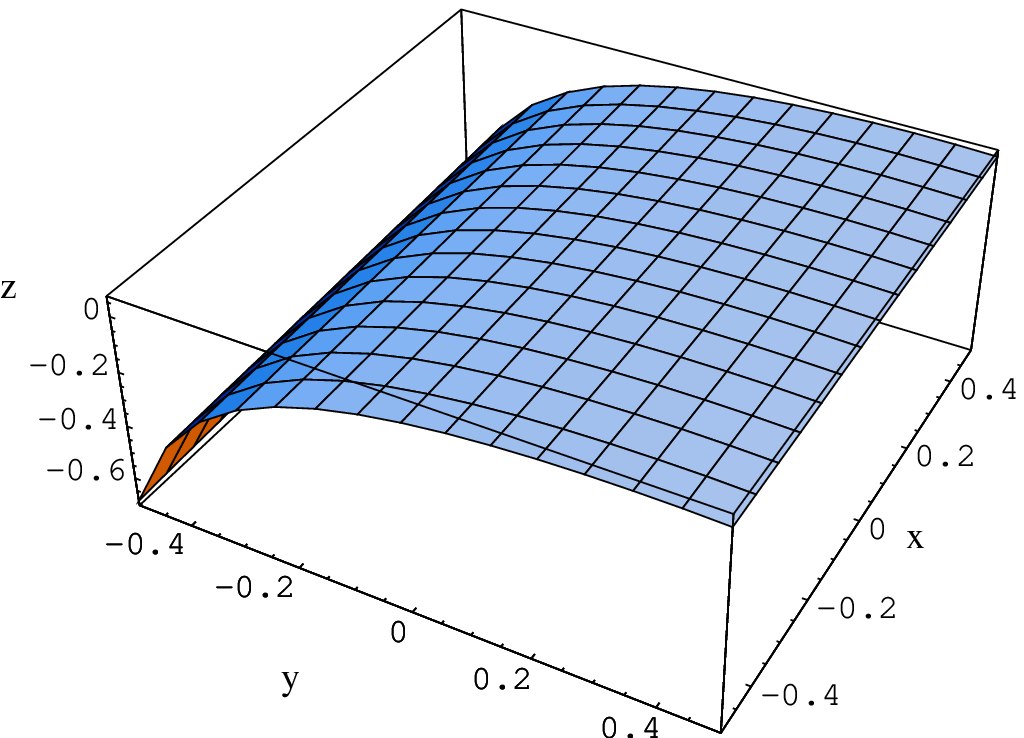}  
      \caption{3D Plot for $R_2$ showing the dependence on 
$\textrm{Re}[\delta]$ and $\textrm{Im}[\delta]$ for the $B_s$ system. }
      \label{bR2MinusDelta3dPlotPhiEqualZero}
    \end{center}
  \end{minipage}
\end{figure}

The effects discussed in this section were calculated using the 
exact expressions for the event rates which are given in the
appendix (section \ref{appendix}) of this article. In these expressions and
for our specific case, where we are interested in $CP$ odd eigenstates,
we set $\eta = 0~\mbox{and}~r=-1$.
In this analysis, we have varied both $\textrm{Re}[\delta]$ and
$\textrm{Im}[\delta]$ in the range between $-0.5$ to $0.5$. 
It has been shown in \cite{Datta:2002wa} that this range is not 
excluded yet by the of the recent Belle data, according to 
which $|m_B - m_{\bar B}|\sim 10^{-14}m_B$ \cite{Leonidopoulos:2001ci}.
 Figs.\ref
{bsR1MinusDelta_2DPlot_RealPart_Streifen} and 
\ref{bR2MinusDelta3dPlotPhiEqualZero} show the results for the $B_s$
system and for appropriate values of the variables. We mainly analyse the
$B_s$ system and also briefly present the results for the $B_d$ system
where one has to account for $CP$ effects arising from mixing.

In Fig. \ref{bsR1MinusDelta_2DPlot_RealPart_Streifen} we show the ratio
$R_1^-$ as a function of $\textrm{Re}[\delta]$ and for three values of
$y$ consistent with the current estimates \cite{lepbosc}.
In the $CPT$ conserving limit, the value for $R_1^-$ can be read off on
the axis $\textrm{Re}[\delta] = 0$ and its range is 
$-0.10 \leq R_1^- \leq -0.025$. When $\textrm{Re}[\delta] \neq 0$ there are
deviations from this range and for 
$\textrm{Re}[\delta] \simeq \pm 0.1$ the deviations for 
$R_1^-$ is $\sim O(0.10)$ from the central line. Thus if a deviation of that
order is observed it will come either from violation of the
$CPT$ symmetry or there is a large width difference of $y\geq 0.12$ either
of which is very interesting. 

Taking $y=0.12$ and $x=19$, in Fig. 
\ref{bR2MinusDelta3dPlotPhiEqualZero} we show the 
ratio $R_2$ as functions
of $\textrm{Re}[\delta]$ and $\textrm{Im}[\delta]$. Again there are sizable
deviations when $\text{Re}[\delta] \neq 0$. The structure of the curves in
 Fig. \ref{bsR1MinusDelta_2DPlot_RealPart_Streifen} and 
Fig. \ref{bR2MinusDelta3dPlotPhiEqualZero} can be accounted for examining
the small $\delta$ limit of the ratios in (\ref{cpt0}). For instance, 
defining $y=\Delta\Gamma/(2\Gamma),~ x=\Delta m/\Gamma$, in the
small $\delta $ limit which we take such that
$(\textrm{Re}[\delta],\textrm{Im}[\delta])\leq 0.1$, we find
\begin{eqnarray}
R_1^-&\approx&\frac{y}{2}+\textrm{Re} \left[\delta\right] \left[\frac{x^2+y^2}
{2\left(x^2+1\right)} \right] + O(\delta ^2)~.
\label{cpt1}
\end{eqnarray}
Thus, a linear dependence of $R_1^-$ as a function of $\textrm{Re}[\delta]$ 
is evident near the origin of the plot in
Fig. \ref{bsR1MinusDelta_2DPlot_RealPart_Streifen}. 
Note that in the limit $y=0$ and 
in the absence of any CP or CPT violation we would expect an equal number of 
$l^+$ and $l^-$ events.

Similarly, we observe in Fig. \ref{bR2MinusDelta3dPlotPhiEqualZero} that for 
small $\textrm{Re}[\delta]$ a linear behavior in $R_2$ and this is again 
understood by examining the $R_2$ in the small $\delta $ limit wherein 
\begin{eqnarray}
R_2&\approx&\frac{1-y}{2(1+x^2)}\Big(\textrm{Re}[\delta]\frac{x^2-y}{2} + 
\textrm{Im}[\delta]x(y+1)\Big)
~ \approx \frac{\textrm{Re}[\delta]}{4}~\mbox{for}~y\ll x~.
\label{cpt2}
\end{eqnarray}
In Fig. \ref{bR2MinusDelta3dPlotPhiEqualZero} for all values of
$\textrm{Re}[\delta]$, the
sensitivity to $\textrm{Im}[\delta]$ is almost negligible. This feature
is shown by the flatness of $R_2$ as a function of $\textrm{Im}[\delta]$. 
In addition without imposing the small $\textrm{Re}[\delta]$ limit, we have 
also numerically checked the dependence of $R_1^-$ on $\textrm{Im}[\delta]$ 
and we find it to be too small $\sim O(0.001)$. This can be attributed to 
the largeness of $x$ for the $B_s$ system which is also brought out in our
approximate analytic expressions discussed above. 

As a consistency check, upon using (\ref{cpt1}) and (\ref{cpt2}) we easily 
find the relation 
\be
R_1^- - 2R_2 = \frac{y}{2}~.
\label{cpt3}
\ee
This relation can be verified by combining
the results in Fig. \ref{bsR1MinusDelta_2DPlot_RealPart_Streifen} and
Fig. \ref{bR2MinusDelta3dPlotPhiEqualZero} where we find 
our numerical results are consistent with (\ref{cpt3}) for the small
$\delta$ limit. Furthermore, as is evident, the ratio $R_2$ is nonzero only 
in the presence of $CPT$ violation and serves as consistency check for 
(\ref{cpt3}). In addition, we note that in the case
of $B_s$ system, for $y\ll x$, in the small $\delta $ limit, we have the 
approximation
\be
R_2\approx R_3 =
\frac{\textrm{Re}[\delta]}{4} + \frac{y\textrm{Im}[\delta]}{2x} 
\approx \frac{\textrm{Re}[\delta]}{4}~.
\label{cpt3a}
\ee 
This calls for the following remark. As a consequence of
the equality, the sum of these two ratios $(R_2 + R_3)$ would prove to be 
a good observable as it does not require time ordering. However, since the 
equality is a result of a specific limit, we may still plot them separately.

 It is interesting to ask what is required in experiments in order to observe
these $CPT$ violating effects. First of all, the effects which are discussed
here are for $C=-1$ state, which is satisfied in $e^+e^-$ colliders; provided
these machines are tuned at higher energies to be able to produce
$B_s\bar B_s$ pairs. A promising 
channel is $B_s \bar B_s \to l^\pm X^\mp J/\psi \phi$ where the lepton and
$J/\psi \phi$ are detected at two different times  with the identification
of the time ordering at the decays, i.e. which decay happened first. The
measurement of the ratios requires an accuracy of $1\%$ and we need
about $10^{4}$ decays to the channels under consideration. 
Given the semi leptonic branching fraction of $B_s$ to be  $10\%$ and the 
decay to $J/\psi\phi$ to be $10^{-3}$ we have a branching fraction for this 
process of $\sim 10^{-4}$. Consequently, we require roughly 
$10^{8}$ $B_s \bar B_s$ pairs produced through a $C=-1$ state. This also
suggests that with the comparable number of events, a reasonable bound
can be set for $\textrm{Re}[\delta]$. Alternatively, if no significant 
deviation from the above mentioned range for say $R_1^-$, is found, then
one can conclude that the data are consistent with zero or rather small
$CPT$ violating effects. Here, by small, we mean $\textrm{Re}[\delta] <
0.1$. This limit can be improved if in future we have better
handle on the values of $y$ which currently suffers from large errors.\\

\begin{figure}[h]
  \begin{minipage}{0.48\textwidth}
    \begin{center}
      \psfrag{z}{$R_1^+$}
      \psfrag{x}{$\textrm{Re}\left[\delta\right]$}
      \psfrag{y}{$\Phi$}
      \includegraphics[width=\columnwidth]{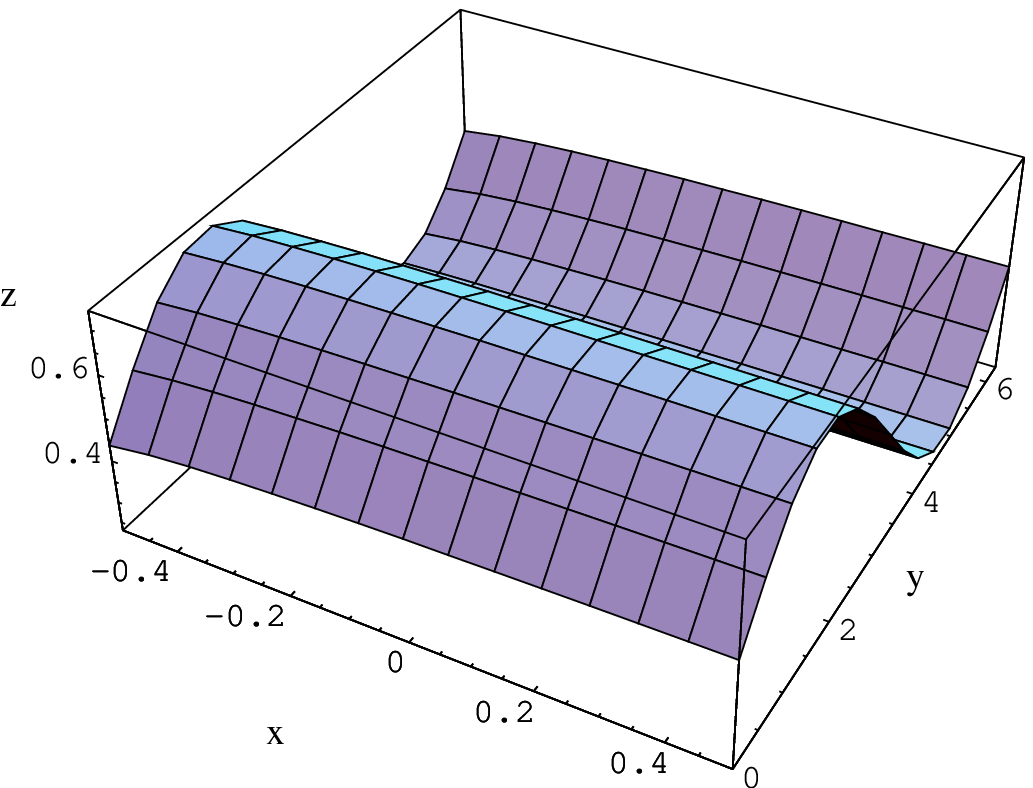}  
      \caption{The ratio $R_1^+$ as a function of $\Phi$ for the $B_d$ 
system. The phase $\Phi$ is given in radians.}
      \label{DependenceOnPhiR1Plus}
    \end{center}
 \end{minipage}
  \begin{minipage}{0.04\textwidth}
     \hfill 
  \end{minipage}%
  \begin{minipage}{0.48\textwidth}
    \begin{center}
      \psfrag{z}{$R_1^-$}
      \psfrag{x}{$\textrm{Re}\left[\delta\right]$}
      \psfrag{y}{$\Phi$}
      \includegraphics[width=\columnwidth]{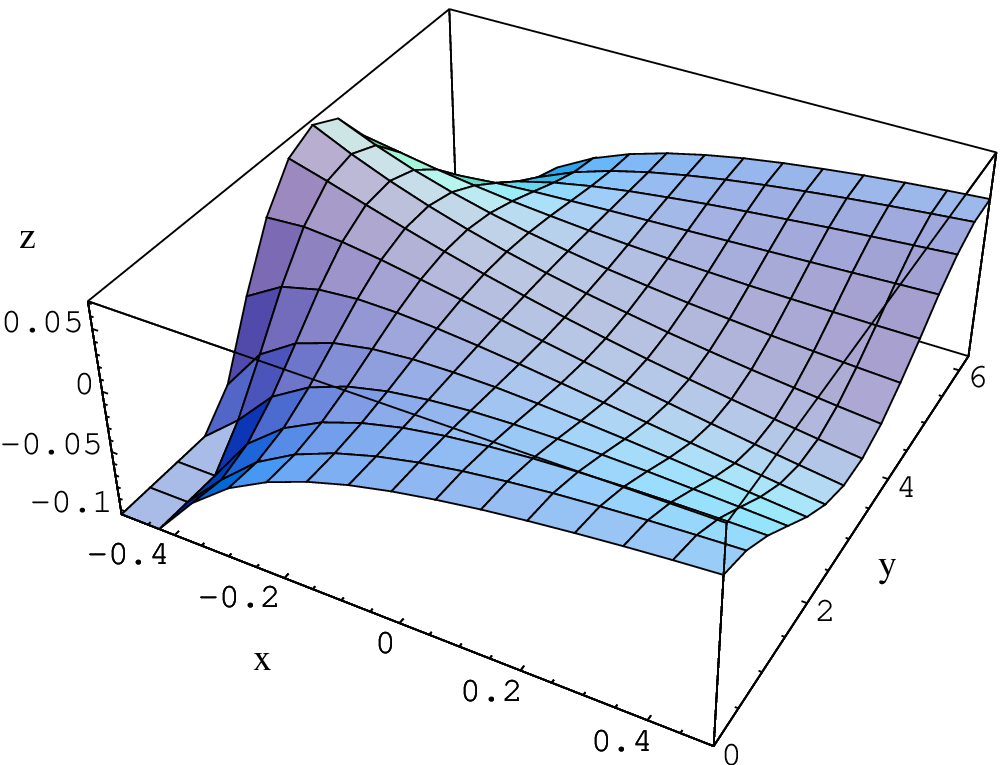}
      \caption{ The same as Fig. \ref{DependenceOnPhiR1Plus} for the ratio
 $R_1^-$.}
      \label{DependenceOnPhiR1Minus}
    \end{center}
  \end{minipage}
\end{figure}

\begin{figure}[h]
  \begin{minipage}{0.48\textwidth}
    \begin{center}
      \psfrag{y}{$R_1^-$}
      \psfrag{x}{$\textrm{Re}\left[\delta\right]$}
      \includegraphics[width=\columnwidth]{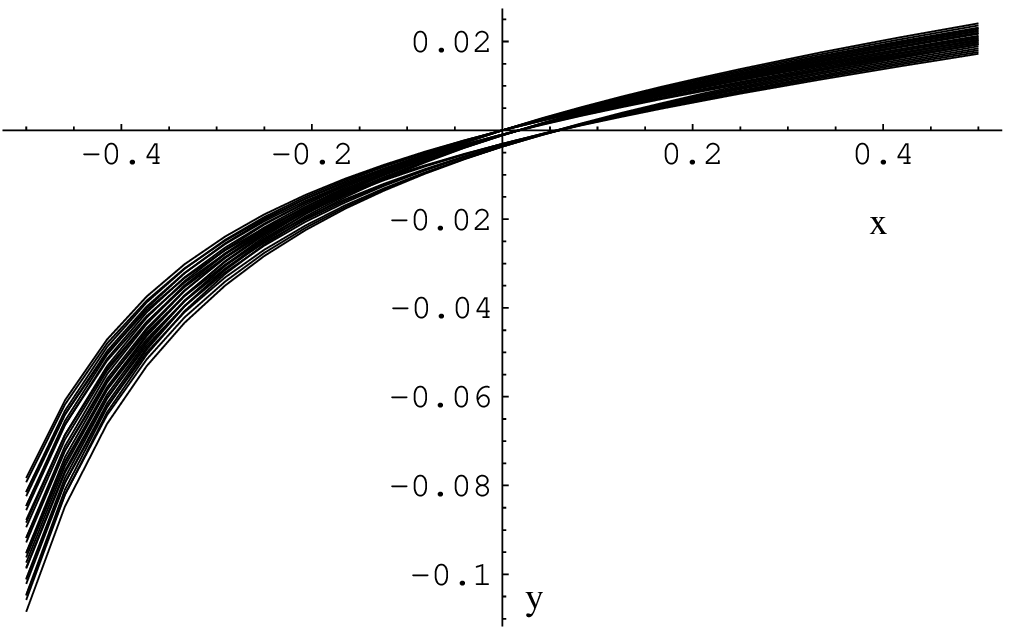}  
      \caption{ Dependence of $R_1^-$ on $\textrm{Re}[\delta]$ 
for the $B_d$ system. The width in the plot is due to the errors in $y,x$ and
$\Phi$. The range of these values are specified in the text.}
      \label{R1MinusDelta_2DPlot_RealPart_Streifen}
    \end{center}
  \end{minipage}
  \begin{minipage}{0.04\textwidth}
     \hfill 
  \end{minipage}%
  \begin{minipage}{0.48\textwidth}
    \begin{center}
      \psfrag{z}{$R_3$}
      \psfrag{x}{$\textrm{Im}\left[\delta\right]$}
       \psfrag{y}{$\textrm{Re}\left[\delta\right]$}
      \includegraphics[width=\columnwidth]{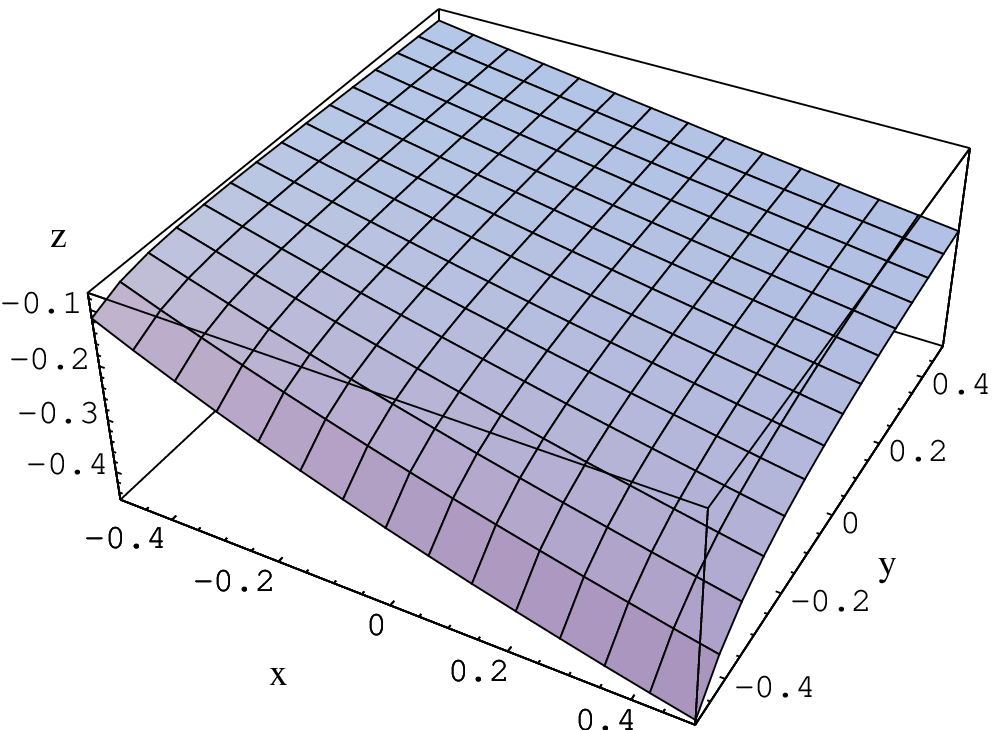}
      \caption{3D Plot for the ratio $R_3$ its dependence on 
$\textrm{Re}[\delta]$ and $\textrm{Im}[\delta]$ for $B_d$ system. 
One can see that the imaginary part is detectable.}
      \label{R3Minus3DPlotPhiEquals2Beta}
    \end{center}
  \end{minipage}
\end{figure}

 We now turn to discuss briefly the effects of $CPT$ violation for the 
$B_d$ system and we choose the process, 
$B_d \bar B_d \to l^\pm X^\mp J/\psi K_S$. In contrast to the $B_s$ mesons, 
for the case of the $B_d$ system, one cannot
neglect the presence of the $CP$ violating phase. We show in Figs.
\ref{DependenceOnPhiR1Plus} and \ref{DependenceOnPhiR1Minus} the variation
of $R_1^\pm$ with $\textrm{Re}[\delta]$ and $\Phi$. In these two plots,
we have set $\textrm{Im}[\delta] =0$ while setting $x=0.755$ and
$y = 10^{-3}$. The variation of 
$R_1^-$ with $\textrm{Re}[\delta]$ is significant and we show this in
Fig. \ref{R1MinusDelta_2DPlot_RealPart_Streifen}. The width in the
plot is due to the allowed range of values in $x$ and $\Phi$. Presently,
their ranges are $\sin\Phi = 0.735 \pm 0.055$ \cite{Pan:2002eg}
with $x = 0.755 \pm 0.015$ and $y \sim 10^{-3}$ \cite{Hagiwara:2002fs}.
The range of $R_1^-$ is now smaller as compared to the $B_s$ case and will
require more events to reach the bound of 
$\textrm{Re}[\delta] \leq \pm 0.1$
Finally, the ratio $R_3$ depends on both 
$\textrm{Re}[\delta]$ and $\textrm{Im}[\delta]$ and illustrated in 
Fig. \ref{R3Minus3DPlotPhiEquals2Beta}. For this plot, we use the values
$\sin\Phi =  0.735$, $x=0.755$ and
$y = 10^{-3}$ . It is interesting that, we can observe the impact of 
$\textrm{Im}[\delta]$ which was not so significant in other plots. In 
contrast to this analysis, the effects of $\textrm{Im}[\delta]$ 
(with $\textrm{Re}[\delta] = 0$) on width measurements and $CP$ asymmetries 
for the $B_d$ mesons has been examined in reference \cite{Datta:2002wa}. 

 Finally, we address the question of nonzero $CP$ violation. 
Although, the present analysis is most effective for the $B_s$ system, we 
note that the ratios in (\ref{cpt0}) can still be used to test for $CPT$ 
violating effects even if the system has a nonzero $CP$ violation. In the pure
$CP$ conserving limit, for instance, independent of $y$, the ratio $R_1^+$ can 
show a numerical deviation from the value of $1/2$ in the presence of $CPT$
violation. This becomes evident by examining the small $\delta$ limit where
\begin{eqnarray}
R_1^+ &\approx& \frac{1}{2} +\textrm{Re}[\delta](\frac{y}{2}) 
- \textrm{Im}[\delta]\frac{x(1-y^2)}{2(1+x^2)}~.
\label{cpt4}
\end{eqnarray}

In the presence of $CP$ violation the ratio $R_1^+$  
gets modified as follows, $R_1^+ \to R_1^+ + r^+$, where the typical 
strength of $r^+$ is a linear combination of the
form $r^+ \sim (\sin\Phi/x, ~\cos\Phi
\cdot \textrm{Re}[\delta]/x)$. Thus, for the $B_s$ system, given the 
expectations for 
$x \sim 19$ and a small $CP$ phase such that $\sin \Phi \sim 10^{-2}$
\cite{bigisanda}, the ratio $R_1^+$ is sensitive to $CPT$ effects 
for all values of 
$(\text{Re}[\delta], \text{Im}[\delta]) \geq r^+ \sim 10^{-3}$.

In the case of CP and CPT conservation we expect the rates $R_S[l^\pm f]= 
R_L[l ^\pm f] = 1/2(R[l^- f]+ R[l^+ f])$ which means that the decays into hadrons 
before the leptonic decay are equal to the hadronic decays which occur after the 
leptonic decay. The equality of the above decay rates is modified in the case that
there is CP or CPT violation.  For $B_d$  decays the $CP$ violation has
been observed and will produce such a difference; an interesting
question here is whether the CP-violation present in $R_1^+$ is
consistent with the CP violation established in other channels,
like $J/\psi K_s$.  Deviations will be attributed to new physics
and/or CPT violation.
Furthermore,
in light of the recent discrepencies in the measurements of $CP$ violation in the 
decays $B_d \to J/\psi K_s$ and  $B_d \to \phi K_s$ one could envisage new physcis 
contribution involving new penguin operators \cite{gudrun}. Such operators could 
induce 
modifications to mixing matrix defined in (\ref{matrix}) and hence modify the 
expected signals from the ratio $R_1^+$. In
the present analysis, we do not consider these possible corrections which may be
required if the present signals for new physics persists and becomes statistically
significant.
 The effects in $B_s$ decays will be even more interesting,
because it is still possible that in extended models  $\sin(\Phi)/x$ is larger than
0.04 and comparison among various channels, like $J/\psi K_s$ with
$\phi K_s$ will be interesting to establish the origin of the effect;
whether the new physics originates in the CP or CPT sector or from both.

 However, we note
that in principle, by strictly measuring $R_1^+$ alone, one cannot ensure the
deviation from $1/2$ as an unambiguous signal for pure $CPT$ violation or pure
$CP$ violation. 
On the other hand, any corrections due to $CP$ can affect 
the ratio $R_1^-$ only at the sub-leading level which is suppressed by 
$(\text{Re}[\delta], \text{Im}[\delta])$ and hence is not significant.
In the present numerical analysis, we consider much larger values 
$(\text{Re}[\delta], \text{Im}[\delta])\sim O(0.1)$ which are not excluded by 
current available data. Clearly, at this level the effects due to $CP$ 
violation are expected to be small for the $B_s$ system. In the case for the $B_d$ 
system, the impact of $\Phi \neq 0$ can be important and must be studied
explicitly in the analysis of the data.

\section{Summary}
\label{summary}
We discussed indirect $CPT$ violation as it appears in the $B$ meson system. 
The break down of the $CPT$ symmetry occurs in several theories and modifies 
the time development of the states. The development is characterized in 
terms of a parameter $\delta$ which is phase-convention independent
\cite{Chou:1999wn}. Consequences of the break down of the $CPT$ symmetries 
manifests
itself in the time development of the states.
 Our attention is addressed to the
production of $B_s\bar B_s$ and $B_d\bar B_d$
pairs in a odd charge conjugation state. In the case of $B_{s}$ and 
$\bar B_{s}$ system, the decay depends on the $CPT$ violating 
parameter $\delta$ while the influence of the $CP$ phase is negligible. 
One way to observe the difference (due to $\delta$) is to study the time 
development of the decays. However, because of the large number of the events
required, we propose time integrated rates. Once the $B_s\bar B_s$ pair is
produced, we encounter two different decays; one of them can be semi-leptonic
and the other one can be hadronic. We also defined a time-ordering among 
them, meaning which one of these events occurs first. 
 
 The time ordered ratio $R_S$ denotes all the
events where the semi-leptonic decay follows the hadronic one; similarly, 
$R_L$ includes the events where the semi-leptonic decay occurs before the
hadronic \cite{Sinha:1998aj}. These two rates are different for two reasons: 
the width differences
and because of $CP~\mbox{and/or}~CPT$ violation. 
 The ratios were calculated
using the formulas given in the appendix and the results were
presented in section \ref{numerical}.
 We calculated the asymmetries involving $R_{L,S}$ as functions of the
parameter $\delta$ and the observed width difference $y$. We summarised our 
numerical results in section \ref{numerical} and through 
Figs. \ref{bsR1MinusDelta_2DPlot_RealPart_Streifen} to 
\ref{R3Minus3DPlotPhiEquals2Beta}. The formalism presented here
holds for $B_d$ system as well and our conclusion is that experiments
with $10^9$ $B_s\bar B_s$ or $B_d \bar B_d$ pairs will be able to restrict
the $\text{Re}[\delta]$ smaller than $10\%$ or otherwise observe an effect.
These considerations are within reach of the present experiments 
\cite{Leonidopoulos:2001ci,aubert}.

It is important to note that 
due to the specific nature of the time ordering, the various 
ratios considered here are sensitive
to $y$ and hence simultaneously to any $CPT$ pollution which can
affect $y$. The impact of $CPT$ violation was shown qualitatively for the 
small $\delta$ limit. In this limit, 
for zero $CP$ violation, which is a good approximation for the $B_s$ system, 
the ratios show a very simple dependence on the parameters $x$ and $y$; thus 
allowing for a direct dependence on $\delta$ under suitable limits 
$(x\gg y)$. We remark that for this limit, the impact of $CPT$ effects 
becomes sensitive to the strength of 
$\textrm{Re}[\delta]~\mbox{and}~\textrm{Im}[\delta]$, more so to 
$\textrm{Re}[\delta]$ due to large $x\gg y$; and thus tests the magnitude and
phase of the $CPT$ violating parameters.
To get a qualitative feeling for the $CPT$ effects, we find from 
(\ref{cpt1}) that for $\textrm{Re}[\delta] \sim y$, one can envisage the
ratios to exhibit 
$CPT$ violating effects $\sim \textrm{Re}[\delta]$ and thus can be
large\footnote{As mentioned earlier,
this effect is observable provided $\textrm{Re}[\delta]$ does not get washed 
out due to the errors in $y$.} $\sim O(0.10)$. In a similar spirit, 
from (\ref{cpt3}) we can expect deviations (here of $O(0.02)$) which is 
relatively independent of the parameters $x$ and $y$. To our knowledge, this 
situation is in contrast with many 
other interesting alternative methods known in the literature where such a 
simple dependence is perhaps not observed in the limit of a small $CPT$ 
violating parameter and thus show a different sensitive to $CPT$ effects 
than the method prescribed here.


\begin{appendix}
\section{Appendix}
\label{appendix}

 In this section, we present exact expressions for the rates calculated
from (\ref{amp}) using the definition

\bea
R_S\left[l^\pm (f, \bar f)\right]&=&\lint_0^\infty dt_0 \lint_0^{t_0} dt 
\abs{A[l^\pm (f, \bar f)]}^2~,\nonumber\\
R_L\left[l^\pm (f, \bar f)\right]&=&
\lint_0^\infty dt_0 \lint_{t_0}^{\infty} dt 
\abs{A[l^\pm (f, \bar f)]}^2~.
\label{rsr}
\eea
 In our notation, $\Omega$ represents the phase of
$\omega \approx 1 + \delta$. Defining the overall prefactor
\be
K = \frac{1+|\omega|^2+2|\omega| \cos \Omega}{2\Gamma^2 |\omega|^2 
\left(1+x^2\right)\left(y^2-1\right)}~,
\label{factor}
\ee
we have:
\begin{eqnarray}
R_S[l^+f]&=&-K\Bigg\{2(x^2+y^2)|\omega|^2|\xi|^2
+2|\xi\omega| r\Big[(1+y)(x^2+y)|\omega|
\cos(\eta+\Phi)\nonumber\\ &-&(y-1)(y-x^2)
\cos(\eta+\Phi-\Omega)+ x(1+y)[|\omega|
  \sin(\eta+\Phi)+\sin(\eta +\Phi-\Omega)]\Big]\nonumber\\
&+ & r^2[(1+x^2)(1+|\omega|^2+y(|\omega|^2-1))+2(y^2-1)|\omega|
(x\sin \Omega-\cos \Omega)]
\Bigg\}~,\nonumber\\
R_L[l^+f]&=& K\Bigg\{
-2(x^2+y^2)|\xi|^2|\omega|^2+r^2(1+x^2)
(-1-y+(-1+y)|\omega|^2)\nonumber\\
&+&
2r|\omega|\Big\{(-1+y^2)\{r\cos\Omega-x|\xi|(|\omega|+
\cos\Omega)\sin(\eta+\Phi)\}\nonumber\\
&+&(1+y)\{rx(-1+y)+(x^2+y)|\xi|  \sin(\eta+\Phi)\}\sin\Omega\nonumber\\
&+&|\xi|\cos(\eta+\Phi)\{(1+y)(x^2+y)
\cos\Omega+(y-1)[(x^2-y)|\omega|+x(1+y)\sin\Omega]\}
\Big\}
\Bigg\}~.\nonumber\\
\end{eqnarray}

\begin{eqnarray}
R_S\left[l^-f\right]&=&K\Bigg\{
-2r^2(x^2+y^2)+(1+x^2)|\xi|^2(-1-y+(y-1)
|\omega|^2)-|\xi|\big[2r\big((1+y)(x^2+y)\nonumber\\ &\times&
\cos(\eta+\Phi)+(y-1)
[(x^2-y)|\omega|\cos(\eta+\Phi-\Omega)+x(1+y)(\sin(\eta+\Phi)\nonumber\\ &+& 
|\omega|\sin(\eta+\Phi-\Omega))]\big)\big]
+2|\xi|^2\left(y^2-1\right)|\omega|\left[\cos\Omega+x\sin\Omega\right]\Bigg\}~,
\nonumber\\
R_L\left[l^-f\right]&=&-K\Bigg\{2r^2[x^2+y^2]+(1+x^2)|\xi|^2(1+|\omega|^2+y
(|\omega|^2-1))-2(y^2-1)|\xi|(|\xi\omega|\cos\Omega\nonumber\\
&+&rx(1+|\omega|\cos\Omega)
\sin(\eta+\Phi)-2(1+y)|\xi\omega|(-x(y-1)\xi+r(x^2+y)\sin(\eta+\Phi))
\nonumber\\ &\times&
\sin\Omega +2r|\xi|\cos(\eta+\Phi)(-(1+y)(x^2+y)\nonumber\\ &\times&
|\omega|\cos\Omega+(y-1)(-x^2+y+x(1+y)|\omega|\sin\Omega))
\Bigg\}~.\nonumber\\
\end{eqnarray}

\begin{eqnarray}
R_S\left[l^+\bar f\right]&=&-K\Bigg\{
2r^2(x^2+y^2)|\xi|^2|\omega|^2
+(1+x^2)(1+|\omega|^2+y(|\omega|^2-1))+2(y^2-1)|\omega|
\nonumber\\ &\times&(x\sin\Omega-\cos\Omega)
+|\omega|\big[
2r|\xi|\big((1+y)(x^2+y)|\omega|\cos(\eta-\Phi)+(y-1)
((x^2-y)\nonumber\\ &\times&\cos(\eta-\Phi+\Omega)
+x(1+y)(|\omega|\sin(\eta-\Phi)+\sin(\eta-\Phi+\Omega)))
\big)\big]\Bigg\}~,
\nonumber\\ 
R_L\left[l^+\bar f\right]&=&K\Bigg\{
-(1+x^2)(1+y)+((1+x^2)(y-1)-2r^2(x^2+y^2)|\xi|^2)|\omega|^2
+2r(x^2-y)\nonumber\\ &\times&(y-1)
|\xi||\omega|^2\cos(\eta-\Phi)+2(y^2-1)|\omega| \cos\Omega
+2|\omega|\big[r(1+y)(x^2+y)|\xi|\nonumber\\ &\times&
\cos(\eta-\Phi+\Omega)
+x(y^2-1)(\sin\Omega+r\xi(|\omega|\sin(\eta-\Phi)+\sin(\eta-\Phi+
\Omega)))\big]\Bigg\}~.\nonumber\\
\end{eqnarray}

\begin{eqnarray}
R_S\left[l^-\bar f\right]&=&K
\Bigg\{-2y^2-r^2|\xi|^2+r^2|\xi|^2(y+|\omega|^2-y|\omega|^2)-x^2(2+
r^2|\xi|^2(1+y+|\omega|^2-y|\omega|^2))\nonumber\\ &-&2r(1+y)(x^2+y^2)
|\xi|\cos(\eta-\Phi)+2r(y-1)|\xi|[r(1+y)|\xi\omega|\nonumber\\ &\times&
  \cos\Omega+(y-x^2)|\omega|\cos(\eta-\Phi+\Omega)
+x(1+y)(\sin(\eta-
\Phi)+r|\xi|\omega\sin\Omega)\nonumber\\ &+&
2rx(y^2-1)|\xi\omega|\sin(\eta-\Phi+\Omega)
\Bigg\}~, 
\nonumber\\ 
R_L\left[l^-\bar f\right]&=&-K\Bigg\{2(x^2+y^2)
+r|\xi|\Big[r|\xi|((1+x^2)(1+|\omega|^2+y(|\omega|^2-1))+2|\omega|(y^2-1)
\nonumber\\
&\times&(-\cos\Omega+x\sin\Omega))+2\Big[(y-1)(y-x^2)\cos(\eta-\Phi)-
(1+y)(x^2+y)\nonumber\\ &\times&|\omega|\cos(\eta-
\Phi+\Omega)
+x(y^2-1)(\sin(\eta-\Phi)+|\omega|(\eta-\Phi+\Omega))\Big]
\Bigg\}~.\nonumber\\
\end{eqnarray}
\end{appendix}

\begin{center} 
{\bf Acknowledgments} 
\end{center} 
This work has been supported by the
Bundesministerium f\"ur Bildung, Wissenschaft, Forschung und Technologie,
Bonn under contract no. 05HT1PEA9. We also like to thank Yuval Grossman,
Amitava Datta and Anirban Kundu for useful comments and discussions.

\newpage



\end{document}